\documentclass[conference]{IEEEtran}
\usepackage[top=0.75in,bottom=1in,left=0.63in,right=0.63in]{geometry}
\IEEEoverridecommandlockouts
\usepackage{url}
\usepackage{amsmath,amsfonts}
\usepackage{graphicx}
\usepackage{cite}
\usepackage{amsmath,amssymb}
\usepackage{booktabs}
\usepackage{multirow}
\usepackage{float}
\usepackage{algorithm}
\usepackage{balance}
\usepackage{algorithmic}
\usepackage{siunitx}
\graphicspath{{images/}}

\begin{document}

%\title{Refined Bayesian Optimization for Efficient Beam Alignment in Intelligent Indoor Wireless Environments}

%\title{Refined Bayesian Optimization for mmWave Beam Alignment: Experimental Results with Commercial 60~GHz Phased Arrays}

\title{Refined Bayesian Optimization for Beam Alignment Based on Real-World 60~GHz Indoor Measurements}

\author{
    \IEEEauthorblockN{
        Parth Ashokbhai Shiroya,
        Amod Ashtekar,
Swarnagowri Shashidhar,
        Mohammed E. Eltayeb
    }
    \IEEEauthorblockA{
        Department of Electrical and Electronics Engineering \\
        California State University, Sacramento \\
        Emails: \{parthshiroya, amodashtekar, sshashidhar, mohammed.eltayeb\}@csus.edu
    }
}

\maketitle

\begin{abstract}
Fast and reliable beam alignment is essential for sustaining directional indoor mmWave links under blockage and multipath. Using recent real-world 60~GHz indoor beam-sweep measurements collected using commercial phased-array transceivers, we find that the measured AoA--AoD power fields are irregular and weakly sparse. The power maps often contain multiple competing lobes and elevated sidelobe floors. These effects stem from dense indoor scattering, sidelobe leakage, and array non-idealities, and violate the sparsity and near-unimodal assumptions often used in beam training.  Motivated by these observations, we propose Refined Bayesian Optimization (R-BO), a power-feedback beam-alignment method that sequentially probes beam pairs to localize the optimum with a limited probe budget. R-BO models the angular power field using a Gaussian Process with a Mat\'ern kernel and selects probes via Expected Improvement. After BO converges, a lightweight one-shot local refinement rescans a small neighborhood around the predicted optimum. Across 43 indoor receiver locations, R-BO achieves 97.7\% exact beam-pair selection accuracy with a beam misalignment penalty below 0.3~dB while reducing probing overhead by 88\% compared to exhaustive beam sweeping.
\end{abstract}

\begin{IEEEkeywords}
Bayesian optimization, Gaussian process, beam alignment,  intelligent wireless environments.
\end{IEEEkeywords}

% ===============================================================
\section{Introduction}
Millimeter-wave (mmWave) and sub-terahertz (sub-THz) communications are central to next-generation wireless systems, enabling multi-Gbps throughput and low latency \cite{xue2024beam,rappaport2019above100}. At these frequencies, high path loss necessitates highly directional beamforming, and links become sensitive to blockage and rapid channel variations, particularly indoors where walls, furniture, and human activity create frequent non-line-of-sight (NLoS) conditions \cite{zhang2023survey,topal2024}. Fast and reliable beam alignment is therefore a key enabler for robust indoor mmWave connectivity.

Beam alignment aims to identify the transmit--receive beam pair that maximizes received power. Exhaustive codebook sweeping guarantees the optimal beam pair, but its overhead scales with the product of the transmit and receive codebook sizes, making it impractical for low-latency operation. To reduce training cost, prior work has developed hierarchical and hybrid training protocols \cite{wang2009beam,alkhateeb2014channel,eltayeb2015opportunistic} and compressive sensing (CS)-based alignment methods \cite{Eltayeb2014,khordad2023compressive}. These approaches typically depend on strong angular sparsity and well-calibrated array responses. In practical indoor deployments, however, dense reflections, sidelobe leakage, off-grid directions, near-field effects, and hardware non-idealities can produce irregular angular power maps with multiple peaks, increasing the likelihood of converging to suboptimal beams. Learning-based beam management using side information has also been explored \cite{wu2022blockage,jiang2023lidar,rinchi2023deep}, but often requires large datasets and nontrivial training and inference complexity.

\textbf{Recent results from real-world measurements:}
This paper is driven by an extensive 60~GHz indoor beam-sweep measurement campaign conducted in a laboratory environment  using commercial phased-array transceivers. The measured AoA--AoD power maps exhibit weak sparsity and pronounced irregularities, including reflection-induced secondary peaks and location-dependent fluctuations, which are often not captured by simulation-only evaluations. These observations motivate a beam-alignment method that operates directly on measured power feedback, adapts online under non-ideal propagation and hardware effects, and remains efficient under a limited probe budget.

\noindent\textbf{Contributions:}
We present a measurement-driven beam-alignment framework based on Refined Bayesian Optimization (R-BO). First, we formulate beam alignment as Bayesian optimization over discrete transmit/receive codebooks using a Gaussian Process surrogate with online hyperparameter re-optimization via log marginal likelihood. Second, we incorporate a lightweight one-shot local refinement stage around the BO-selected beam pair to improve robustness on measured indoor power maps. Third, using real 60~GHz measurements across 43 receiver locations, we demonstrate near full-sweep performance with an 88\% reduction in probing overhead relative to exhaustive sweeping.

\section{System Model}
We consider an indoor mmWave beamforming link between a transmitter (TX) with $t$ antennas and a receiver (RX) with $r$ antennas. The TX and RX employ predefined beamforming codebooks
$\mathbf{F}=\{\mathbf{f}_1,\ldots,\mathbf{f}_p\}$ and $\mathbf{W}=\{\mathbf{w}_1,\ldots,\mathbf{w}_q\}$,
where $\mathbf{f}_i\in\mathbb{C}^{t\times 1}$ and $\mathbf{w}_j\in\mathbb{C}^{r\times 1}$ denote the transmit and receive beamforming vectors, respectively. Following the standard narrowband mmWave MIMO model \cite{alkhateeb2014channel}, the received signal for beam pair $(i,j)$ is
\begin{equation}
y_{i,j}=\mathbf{w}_j^{H}\mathbf{H}\mathbf{f}_i s + e,
\end{equation}
where $\mathbf{H}\in\mathbb{C}^{r\times t}$ is the channel matrix, $s$ is a unit-power pilot symbol with $\mathbb{E}[|s|^2]=1$, and $e\sim\mathcal{CN}(0,\sigma^2)$ denotes additive white Gaussian noise.

%Classical mmWave models often assume angular sparsity, i.e., only a few dominant propagation paths contribute most of the received energy. However, in practical indoor environments, dense scattering, near-field effects, sidelobe leakage, and transceiver non-idealities can yield irregular angular power profiles that are not strongly sparse. Consequently, the received power may be distributed across multiple candidate beams rather than concentrated in a small set of AoD/AoA pairs. This motivates beam-alignment strategies that can efficiently identify high-power beam directions using limited measurements.

For a beam pair $(i,j)$, we define the received power metric as
\begin{equation}
P_{i,j} \triangleq \left|\mathbf{w}_j^{H}\mathbf{H}\mathbf{f}_i\right|^2.
\end{equation}
The goal of beam alignment is to identify
\begin{equation}
(i^\star,j^\star) \in \arg\max_{1\le i \le p,\; 1\le j \le q} \; P_{i,j},
\end{equation}
while minimizing the number of probed beam pairs.

\section{Proposed Bayesian Optimization Framework for Beam Alignment}
This section presents Refined Bayesian Optimization,  a measurement-driven beam-alignment method that operates directly on received-power feedback. R-BO formulates beam alignment as a sequential learning problem that exploits the structured angular power distribution inherent in directional transceivers.  In practical indoor mmWave deployments, dense reflections and hardware non-idealities create non-convex AoA--AoD power fields characterized by multiple lobes, e.g.  one dominant mainlobe and several weaker sidelobes and reflection-driven secondary peaks. As the transmitter and receiver steer their beams, sidelobe energy ``leaks'' into neighboring angles, providing informative local structure that can indicate the direction of the true optimum even before it is directly probed. R-BO leverages this angular leakage together with model uncertainty to efficiently navigate the irregular angular landscape.

Rather than relying on strong angular sparsity or idealized array responses, R-BO treats beam alignment as a sequential decision problem over discrete TX/RX codebooks and allocates probes to beam pairs that are most informative under the current surrogate model. By balancing exploration of uncertain regions against exploitation of high-power candidates, the algorithm progressively focuses the search toward the strongest beam pair with a limited number of measurements.

\begin{figure}[!t]
    \centering
    \includegraphics[width=\linewidth]{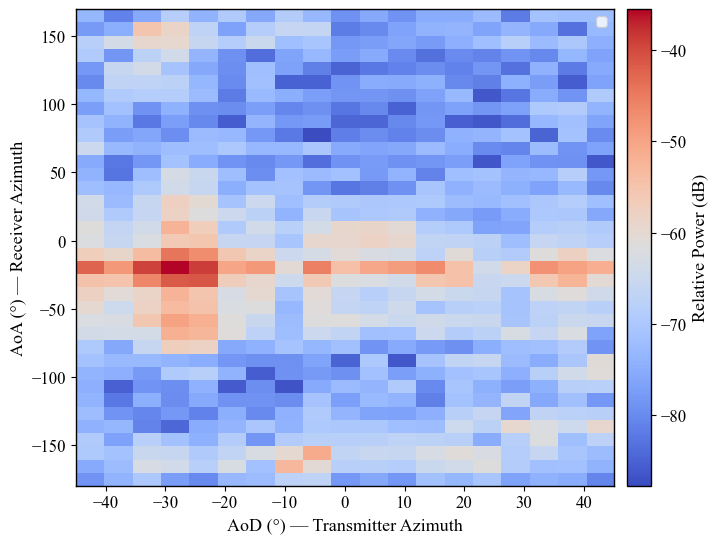}
    \caption{Measured AoA--AoD relative power map under exhaustive sweeping (RX23). Measurements use 16 TX and 16 RX antennas at 60~GHz in an indoor laboratory.}
    \label{fig:rx23_truth2}
\end{figure}

Fig.~\ref{fig:rx23_truth2} shows a representative measured AoA--AoD relative power map from an indoor laboratory measurement (location ID RX23) obtained via exhaustive sweeping. The irregular lobes and reflection-driven peaks motivate a data-driven strategy that balances exploration and exploitation under a limited probe budget. R-BO proceeds in three stages: (i) a small random initialization to fit an initial surrogate model, (ii) iterative Bayesian optimization to adaptively select beam probes, and (iii) a lightweight one-shot refinement around the predicted optimum.

\subsection{Gaussian Process Surrogate}
R-BO models the unknown angular power field as a Gaussian Process (GP),
\begin{equation}
f(\mathbf{x}) \sim \mathcal{GP}\big(\mu(\mathbf{x}), k(\mathbf{x}, \mathbf{x}')\big),
\end{equation}
where $\mathbf{x}$ denotes the beam-pair coordinates (AoD, AoA). We adopt a zero-mean prior and encode AoA/AoD using sine--cosine features to preserve angular periodicity and avoid discontinuities at $\pm180^\circ$. A Mat\'ern kernel captures moderate local smoothness across neighboring beams, which is consistent with practical beam-pattern continuity and sidelobe structure. To accommodate location-dependent and non-stationary indoor power maps, kernel hyperparameters are re-optimized online via marginal-likelihood maximization as new measurements are acquired.

\begin{figure}[!t]
    \centering
    \includegraphics[width=\linewidth]{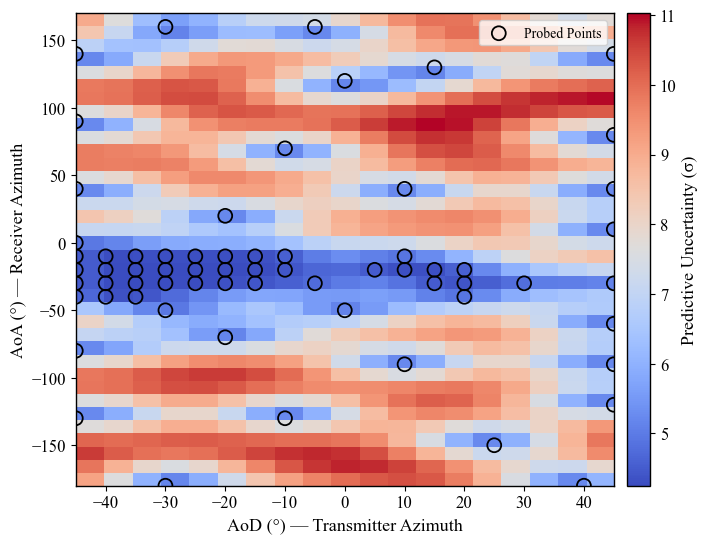}
    \caption{GP posterior uncertainty map for RX23 after BO (before refinement). Black circles indicate probed beam pairs.}
    \label{fig:gp_uncertainty}
\end{figure}

Fig.~\ref{fig:gp_uncertainty} illustrates the GP posterior standard deviation for RX23 after the BO phase (before refinement). Unprobed regions exhibit higher uncertainty, while sampled regions show lower uncertainty, enabling uncertainty-aware probing.

\subsection{Acquisition and Sampling Strategy}
At iteration $t$, R-BO selects the next beam pair by maximizing the Expected Improvement (EI) acquisition function,
\begin{equation}
\mathrm{EI}(\mathbf{x}) = (\mu_t - f_t^{+} - \xi)\Phi(Z_t) + \sigma_t\phi(Z_t),
\end{equation}
where $Z_t = (\mu_t - f_t^{+} - \xi)/\sigma_t$, $f_t^{+}$ is the best measured power so far, and $\Phi(\cdot)$ and $\phi(\cdot)$ are the standard normal CDF and PDF. EI is well-suited to measured indoor power fields because it trades off exploitation of high predicted power against exploration in high-uncertainty regions. We set $\xi=0.05$ to encourage broader exploration early and stronger exploitation as the surrogate becomes more accurate.

\subsection{One-shot Refinement}
After $T$ BO iterations, we apply a one-time local refinement centered at the beam pair that maximizes the GP posterior mean. Rather than performing an iterative local optimizer, refinement simply rescans a small fixed neighborhood and selects the beam pair with the highest measured power within that set. In our implementation, refinement evaluates 15 neighboring beams around the BO-selected TX/RX indecies yielding a constant refinement cost of $N_{\text{ref}}=15$ additional probes. This step adds negligible overhead and mitigates discretization and surrogate mismatch without retraining the GP. Algorithm~\ref{alg:rbo2} summarizes the complete procedure. Unless stated otherwise, we use $n_{\text{init}}=15$, $T=50$, and a $\pm10^\circ$ refinement neighborhood, with early stopping when $\max_{\mathbf{x}}\mathrm{EI}(\mathbf{x})<10^{-8}$. The GP surrogate is implemented using \texttt{scikit-learn}'s \texttt{GaussianProcessRegressor} with a Mat\'ern kernel and a white-noise term.

\begin{figure}[!t]
    \centering
    \includegraphics[width=\linewidth]{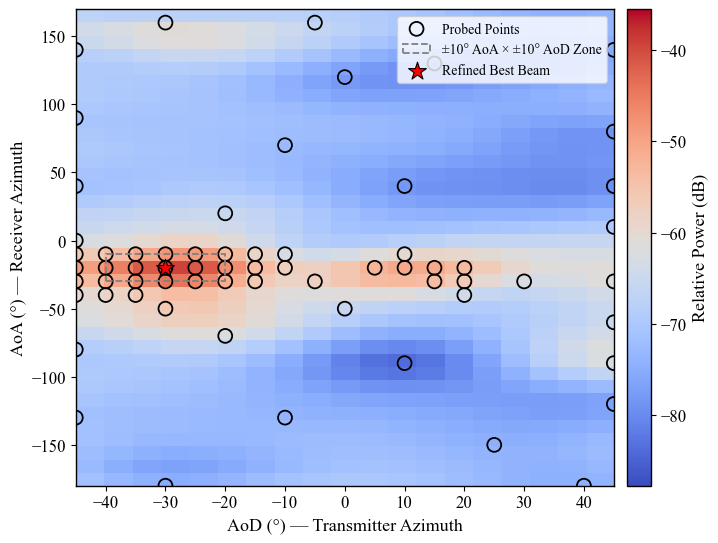}
    \caption{GP posterior mean for RX23 after R-BO. The red star marks the final beam; the dashed box indicates the $\pm10^{\circ}$ refinement neighborhood.}
    \label{fig:rx23_bo_mean}
\end{figure}

\begin{algorithm}[!t]
\caption{R-BO algorithm for beam alignment.}
\label{alg:rbo2}
\begin{algorithmic}[1]
\REQUIRE Beam space $\mathcal{X}$, initial probes $n_{\text{init}}$, iterations $T$, exploration constant $\xi$
\ENSURE Estimated optimal beam pair $\mathbf{x}_{\text{ref}}^{\star}$
\STATE Randomly probe $n_{\text{init}}$ beam pairs $\{\mathbf{x}_i\}$ and measure powers $\{y_i\}$.
\STATE Fit a GP with a Mat\'ern kernel to $(\mathbf{X},\mathbf{y})$.
\FOR{$t = 1$ to $T$}
    \STATE Compute $\mu_t(\mathbf{x})$ and $\sigma_t(\mathbf{x})$ for unprobed beams.
    \STATE Select $\mathbf{x}_{t+1}=\arg\max_{\mathbf{x}}\mathrm{EI}(\mathbf{x})$ and measure $y_{t+1}$.
    \STATE Append $(\mathbf{x}_{t+1},y_{t+1})$ and refit the GP.
    \STATE Re-optimize kernel hyperparameters via log marginal likelihood.
\ENDFOR
\STATE \textbf{Refinement:} Rescan a fixed neighborhood around $\arg\max_{\mathbf{x}}\mu_T(\mathbf{x})$ by probing  neighboring TX/RX beams.
\RETURN $\mathbf{x}_{\text{ref}}^{\star}=\arg\max_{\mathbf{x}} y(\mathbf{x})$.
\end{algorithmic}
\end{algorithm}

\section{Data Collection and Experimental Setup}
To evaluate beam alignment under practical indoor propagation and hardware effects, we conducted a real-world 60~GHz beam-sweep measurement campaign in the Wireless Systems Laboratory at California State University, Sacramento. The objective was to obtain measured AoD--AoA power maps using commercial phased-array transceivers in a cluttered indoor environment, thereby capturing effects that are often absent from simulation-only studies.

\begin{figure}[t]
    \centering
    \includegraphics[width=0.9\linewidth]{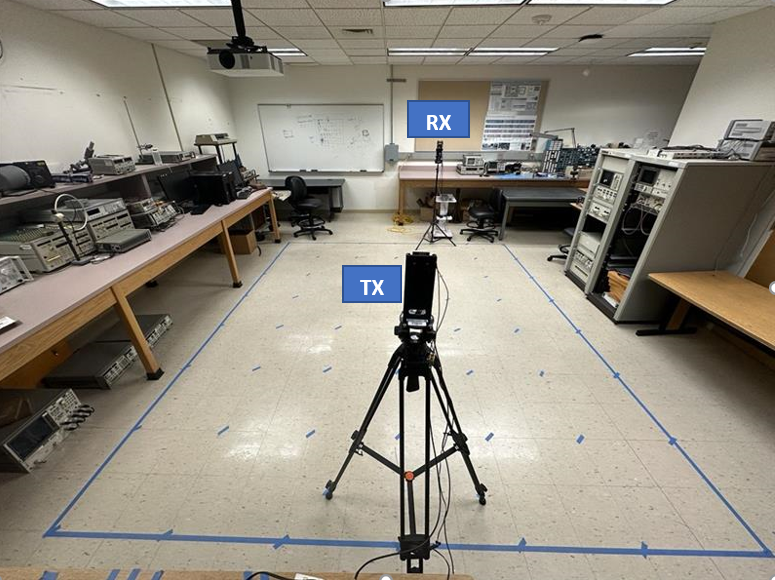}
    \caption{Room geometry and experimental setup in the laboratory. The TX was fixed at a single location and oriented toward the far wall, while the RX was moved sequentially across marked grid points to capture spatially distributed beam-sweep measurements.}
    \label{fig:room_geometry}
\end{figure}

Measurements were collected in a closed $12~\text{m} \times 8~\text{m}$ laboratory containing metallic benches, computers, and wooden tables, producing rich multipath scattering representative of indoor mmWave deployments. The transmitter and receiver were mounted on tripods at a height of $1.6~\text{m}$ to maintain approximate coplanarity and emulate a typical access-point (AP) to user-equipment (UE) geometry, as illustrated in Fig.~\ref{fig:room_geometry}. The floor area was discretized into a $7\times8$ grid (56 marked points). The TX remained fixed at a single location and orientation, while the RX was moved sequentially across the remaining grid points. We report results over 43 RX locations selected to ensure stable measurements and representative spatial coverage of the room.

The measurement platform consisted of a pair of Sivers Semiconductors EVK06002 phased-array transceivers operating at 60~GHz in analog beamforming mode. At the TX, electronic beam steering swept from $-45^{\circ}$ to $+45^{\circ}$ in $5^{\circ}$ increments, yielding 19 transmit beams. At the RX, a motorized gimbal enabled mechanical scanning from $-180^{\circ}$ to $180^{\circ}$ in $10^{\circ}$ increments, yielding 36 receive directions. Each RX location therefore produced $19\times36=684$ AoD--AoA beam-pair power measurements, which serve as the exhaustive beam-sweep baseline in Section~\ref{sec:results}. Transmit power and array configuration parameters (including gain settings) were held constant throughout the campaign to ensure consistent comparisons across locations and sweeps.

\section{Experimental Results and Discussions}
\label{sec:results}
This section reports recent experimental results for the proposed Refined Bayesian Optimization (R-BO) framework using the measured 60~GHz indoor dataset collected in a laboratory environment. Results are averaged over 43 receiver  locations with direct line-of-sight (LoS) to the transmitter. The goal is to quantify how closely R-BO approaches exhaustive beam sweeping on real measured AoA--AoD power maps, while substantially reducing the number of required beam probes.

\subsection{Evaluation Metrics}
Performance is evaluated using three metrics reported throughout this section. Beam-alignment accuracy is the percentage of RX locations for which the final beam pair selected by R-BO (after refinement) exactly matches the exhaustive-search optimum. The beam misalignment penalty (dB) is the received-power gap between the beam pair selected by an algorithm and the exhaustive-search best beam pair at the same RX location, averaged across RX locations; it directly quantifies the link-performance loss due to selecting a suboptimal beam. Probing overhead is the total number of beam-pair probes per RX location, including initialization, BO iterations, and refinement, and thus captures the beam training overhead.

Unless stated otherwise, R-BO uses a one-shot refinement radius of $\Delta=\pm10^\circ$, applied only after the BO phase. Importantly, $\Delta$ is not an accuracy tolerance; an RX location is counted as correct only when the final selected beam pair exactly equals the exhaustive-search optimum. The total probe budget per RX location is
$
N_{\text{tot}} = n_{\text{init}} + T + N_{\text{ref}},
$
where $n_{\text{init}}$ is the number of random initialization probes and $T$ is the number of BO iterations. Refinement is executed once around the best BO-selected pair. Unless stated otherwise, we use $n_{\text{init}}=15$ and $T=50$, resulting in $N_{\text{tot}}=80$ probes per RX location.

\begin{figure}[!t]
    \centering
    \includegraphics[width=\linewidth]{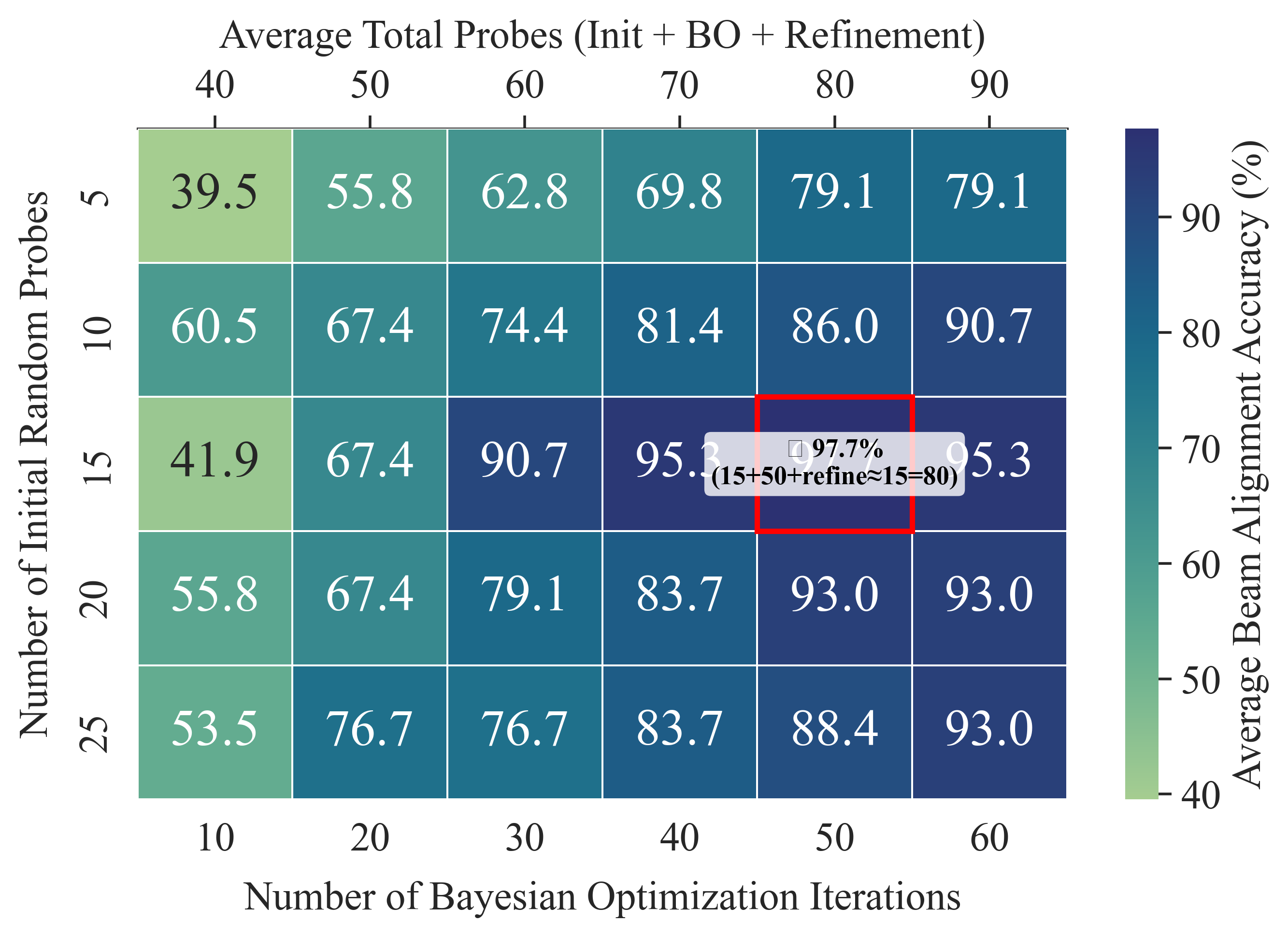}
    \caption{Beam-alignment accuracy (\%) as a function of initialization size and Bayesian iterations. The red cell marks ($n_{\text{init}}{=}15$, $T{=}50$), which provides the best accuracy--efficiency trade-off on the measured dataset.}
    \label{fig:heatmap}
\end{figure}

\subsection{Sensitivity to Initialization and Iteration Budget}
Fig.~\ref{fig:heatmap} shows beam-alignment accuracy as a function of the initialization size $n_{\text{init}}$ and the BO iteration budget $T$. The figure highlights the trade-off between prior sampling and adaptive exploration on measured indoor power maps. With small initialization sets ($n_{\text{init}}<10$), the GP surrogate is under-trained and EI can be misled by local irregularities and secondary peaks, resulting in lower accuracy. Increasing $n_{\text{init}}$ improves early model reliability; however, overly large initializations reduce sample efficiency because many probes are spent before uncertainty-guided sampling begins.

A similar trend is observed for $T$, the number of BO iterations after the $n_{\text{init}}$ random probes. With small iteration budgets, the algorithm may not sufficiently explore the angular space and can miss reflection-driven secondary peaks, leading to suboptimal beam selection. As $T$ increases, BO typically localizes the dominant high-power region and then concentrates subsequent probes in that neighborhood; consequently, very large $T$ yields diminishing returns because additional samples rarely change the best discrete beam on the codebook. In our dataset, the best accuracy--efficiency trade-off occurs at $(n_{\text{init}},T)=(15,50)$, which we adopt for the remainder of the experiments. Beyond $N_{\text{tot}}= $80 probes,  performance improvements fluctuate and saturate, consistent with diminishing marginal gains once the dominant region has been identified on the discrete 10$^\circ$ beam grid.

\begin{figure}[!t]
    \centering
    \includegraphics[width=0.9\linewidth]{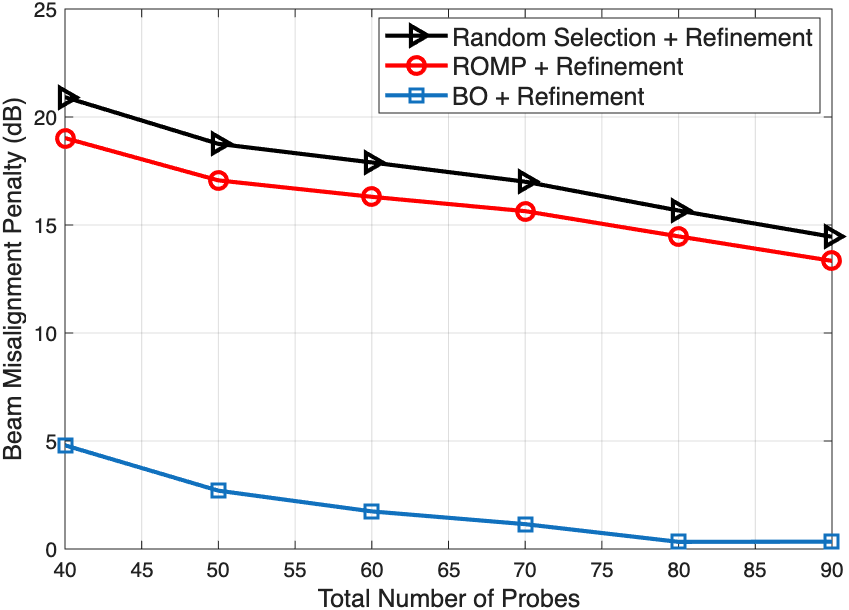}
    \caption{Average beam misalignment penalty (dB) versus probe count $N_{\text{tot}} $ for R-BO, ROMP \cite{Needell2010ROMP}, and random probing under the same total probe budget (including refinement). R-BO converges faster and achieves a lower steady-state penalty on measured indoor data.}
    \label{fig:power_loss}
\end{figure}

\subsection{Convergence Behavior and Baseline Comparisons}
Fig.~\ref{fig:power_loss} plots the beam misalignment penalty versus the total number of probes $N_{\text{tot}} $. The probe count includes $n_{\text{init}}$ initialization probes, $T$ BO iterations, and (when enabled) the $N_{\text{ref}}=15$ refinement probes. R-BO reduces the misalignment penalty sharply within the first 20--40 probes, indicating that EI rapidly steers sampling toward informative high-power angular regions in the measured maps. Beyond $\sim$80 total probes, the penalty stabilizes below 0.3~dB, suggesting that R-BO consistently identifies beams near the exhaustive optimum across most RX locations while avoiding redundant probing.

We compare R-BO against Regularized Orthogonal Matching Pursuit (ROMP) \cite{Needell2010ROMP} and random beam probing under the same total probe budget, and we apply the same one-shot refinement step to all methods for fairness. ROMP relies on angular sparsity and ideal array responses; in measured indoor settings with reflections, sidelobe leakage, and hardware non-idealities, these assumptions are weakened. Consequently, ROMP converges more slowly and exhibits a higher residual misalignment penalty than R-BO. Random probing performs comparably to ROMP in this dataset, indicating that sparsity-driven recovery provides limited benefit when the measured angular power field is weakly sparse and irregular.

\begin{figure}[!t]
    \centering
    \includegraphics[width=0.9\linewidth]{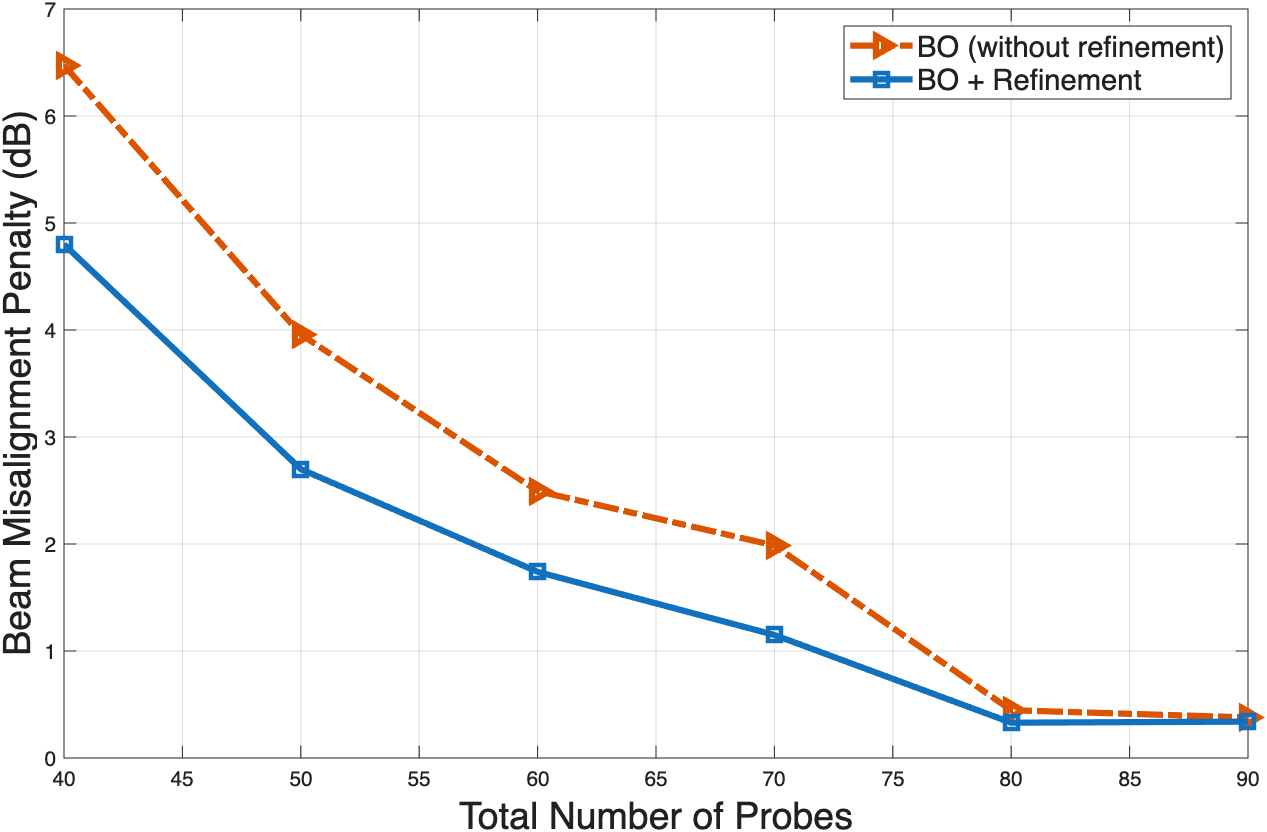}
    \caption{Average beam misalignment penalty (dB) versus probe count ($n_{\text{init}}{=}15$) for Bayesian optimization with and without refinement. Refinement improves performance at low probe budgets; both approaches converge as the probe budget increases.}
    \label{fig:power_loss2}
\end{figure}

\subsection{Effect of the One-shot Refinement}
Fig.~\ref{fig:power_loss2} compares Bayesian optimization with and without the proposed refinement stage. Refinement provides the largest benefit at low-to-moderate probe budgets by correcting near-misses, i.e., cases where BO identifies a neighborhood around the optimum but does not directly probe the globally optimal discrete beam pair. As the probe budget increases, BO without refinement increasingly samples near the optimum and both curves converge to the same floor. This behavior is expected for a discrete 10$^\circ$ grid and once BO localizes the optimum within one codebook step, the refinement probes become largely redundant and rarely change the final selected beam pair.

\begin{figure}[!t]
    \centering
    \includegraphics[width=0.9\linewidth]{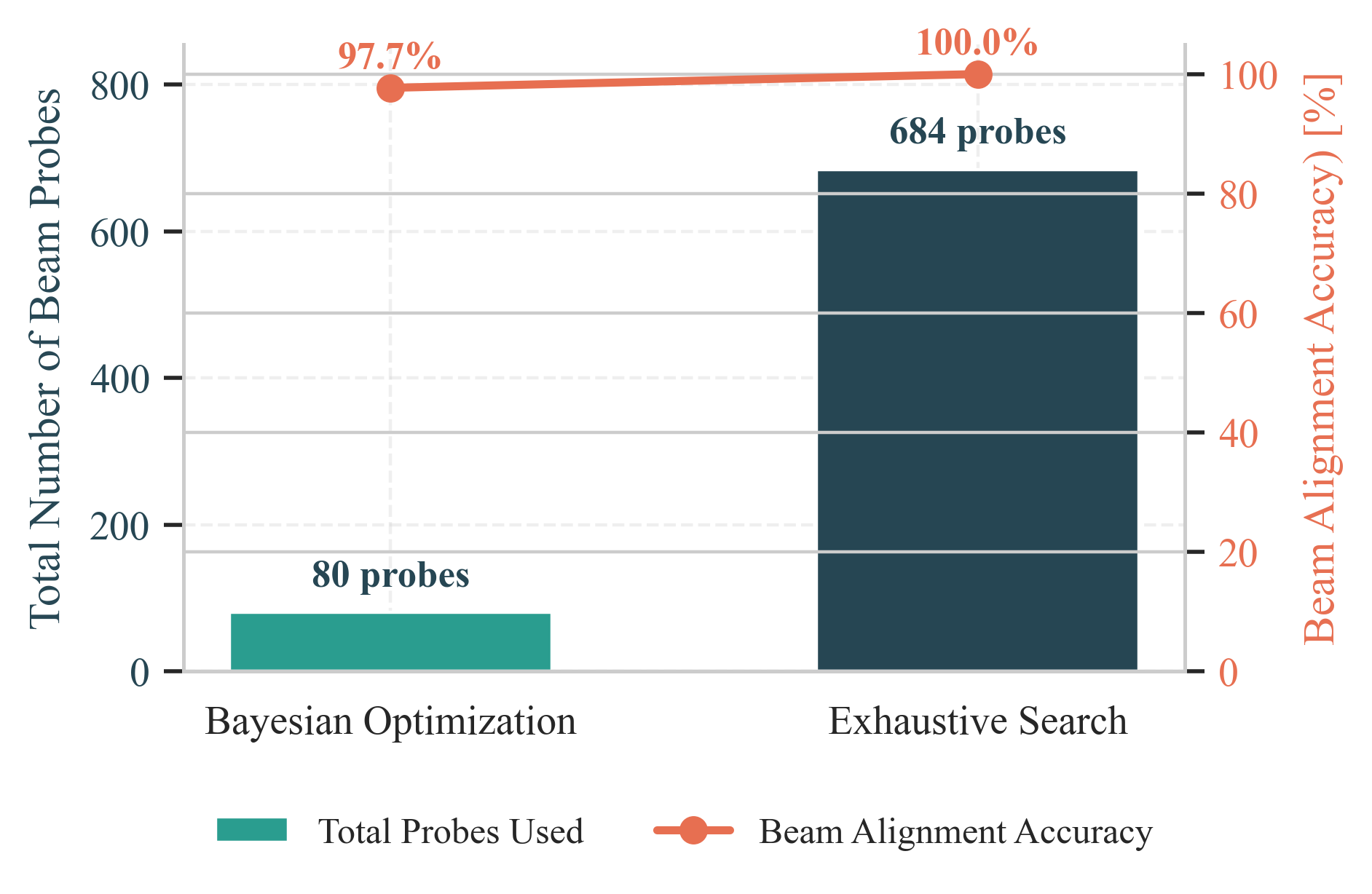}
    \caption{Beam-alignment performance comparison between R-BO and exhaustive beam sweeping across 43 RX locations. R-BO approaches full-sweep accuracy with an 88\% reduction in probing overhead.}
    \label{fig:bo_vs_exhaustive}
\end{figure}

\subsection{Overall Performance vs.\ Exhaustive Beam Sweep}
Fig.~\ref{fig:bo_vs_exhaustive} summarizes overall performance relative to exhaustive beam sweeping. Exhaustive sweeping tests all 684 TX--RX beam pairs per RX location and thus achieves perfect alignment at prohibitive measurement cost. In contrast, R-BO achieves 97.7\% exact beam-pair selection accuracy while requiring only $N_{\text{tot}}=80$ probes, corresponding to an 88\% reduction in probing overhead. The misalignment-penalty trends in Figs.~\ref{fig:power_loss} and \ref{fig:power_loss2} further show that R-BO attains near-exhaustive performance under a limited probe budget, supporting low-latency beam training in practical indoor deployments.

%\subsection{Discussion}
%Overall, the results indicate that the measured indoor angular power field contains exploitable structure---including smooth local variation and sidelobe correlations---that a GP surrogate can capture even when strict angular sparsity does not hold. R-BO benefits from principled exploration in uncertain regions, exploitation of dominant lobes guided by EI, and a final local scan that mitigates discretization and surrogate mismatch. The saturation of refinement gains at high probe budgets is therefore expected: once BO consistently probes within one beam of the optimum on the discrete 10$^\circ$ grid, a one-shot neighborhood scan with a fixed number of refinement probes rarely yields additional improvement.

%%%%%MEMEME

\section{Conclusion}
This paper presented Refined Bayesian Optimization for efficient beam alignment in practical indoor mmWave deployments. The study is motivated and validated using real 60~GHz indoor beam-sweep measurements collected in a laboratory with commercial phased-array transceivers. The measured AoA--AoD power maps are often weakly sparse and exhibit sidelobe leakage and reflection-induced secondary peaks, which can reduce the benefit of sparsity-driven power-based beam training. R-BO aligns beams directly from measured power feedback using a Gaussian Process surrogate, Expected-Improvement-guided probing, and a lightweight one-shot local refinement. Across 43 LoS receiver locations, R-BO achieves 97.7\% exact beam-pair selection accuracy with a beam misalignment penalty below 0.3~dB while reducing probing overhead by 88\% relative to exhaustive sweeping. Compared with sparsity-driven baselines such as ROMP, R-BO converges faster and is more robust to measured non-ideal propagation and hardware effects, supporting low-latency indoor beam training. Future work will extend this work to online beam tracking under mobility and explore integrating side information (e.g., LiDAR) to further improve adaptation across diverse indoor scenarios.

\section*{Data Availability}
The beam-sweep measurement dataset supporting the findings of this paper has been made publicly available at:
\url{https://github.com/72784/MMWAVE-Dataset}

\section*{Acknowledgment}
This work was supported by the National Science Foundation under Grant No.~NSF-2243089.

\balance

\end{document}